\documentclass[
aps,%
12pt,%
final,%
notitlepage,%
oneside,%
onecolumn,%
nobibnotes,%
nofootinbib,
superscriptaddress,%
noshowpacs,%
centertags]%
{revtex4}
\RequirePackage{graphicx}

\def\refitem#1{\relax}

\begin{document}
\title{The Chiral Magnetic Effect: Beam-energy and system-size dependence }

\author{\firstname{V.D.} \surname{Toneev}}
\email{toneev@theor.jinr.ru}
\affiliation{Bogoliubov  Laboratory of Theoretical Physics, JINR Dubna, 
141980 Dubna, Russia}
\affiliation{GSI, Helmholtzzentrum f\"ur
Schwerionenforschung GmbH, 64291 Darmstadt, Germany}

\author{\firstname{V.} \surname{Voronyuk}}
\email{vadimv@jinr.ru}
\affiliation{Laboratory of High Energy Physics, JINR, 
141980 Dubna, Russia}
\affiliation{Bogolyubov
Institute for Theoretical Physics, 03680 Kiev, Ukraine}
\affiliation{Institute
f\"ur Theoretische Physik, Universit\"at Frankfurt, 60438 Germany}

	\begin{abstract}
We consider the energy dependence of the local ${\cal P}$ and
${\cal CP}$ violation in Au+Au and Cu+Cu collisions over a large energy
range within a simple phenomenological model. It is
expected that at LHC the chiral magnetic effect   will be   about
20 times weaker than at RHIC. At lower energy range  
this effect should vanish sharply at energy somewhere above
the top SPS one. To elucidate CME background effects a transport model
including magnetic field evolution is put forward. 
\end{abstract}

\maketitle

\def\lsim{\mathrel{\rlap{
\lower4pt\hbox{\hskip-3pt$\sim$}}
\raise1pt\hbox{$<$}}}     
\def\gsim{\mathrel{\rlap{
\lower4pt\hbox{\hskip-3pt$\sim$}}
\raise1pt\hbox{$>$}}}     
\def\be{\begin{eqnarray}}
\def\ee{\end{eqnarray}}
\def\prt{\partial}

\section{\label{intro}Introduction}
As was
argued in Refs.~\cite{Kharzeev:2004ey,KZ07,KMcLW07,FKW08} the QCD
topological effects in intimate connection with axial anomaly may 
be observed in heavy ion collisions
directly in the presence of very intense external electromagnetic
fields due to the ``Chiral Magnetic Effect'' (CME) as a
manifestation of spontaneous violation of the ${\cal CP}$
symmetry.  First
experimental evidence for the CME identified via the observed
charge separation effect with respect to the reaction plane has been
presented by the STAR Collaboration at RHIC~\cite{Vo09}. In this
paper we analyze the STAR data in a simple phenomenological way to
estimate a possibility observing the CME in the larger energy
range, from the LHC to FAIR/NICA energies. We also make a step towards
a dynamical estimate of the CME background based on the nonequilibrium
Hadron-String-Dynamics (HSD) microscopical transport approach \cite{HSD}
including magnetic field.

\section{Phenomenological estimates of the CME}\label{sec:2}

A characteristic scale of
the process is given by the saturation momentum 
$Q_s$~\cite{Kharzeev:2004ey}, so the
transverse momentum of particles $p_t \sim Q_s$. Then the total transverse
energy per unit rapidity at mid-rapidity deposited at
the formation of hot matter is expressed through the  overlapping surface
of two colliding nuclei in the transverse plane $S$
\begin{eqnarray}
\label{energy}
\frac{dE_T}{dy} &\sim& \epsilon \cdot V=\epsilon \cdot \Delta z \cdot S
= Q_s\cdot (Q_s^2 S)  \sim Q_s \cdot \frac{dN_{\rm hadrons}}{dy}~.
\end{eqnarray}
Here the energy density and longitudinal size $\Delta z \simeq \Delta \tau
\simeq 1/Q_s$ are taken in order of magnitude as follows
$\epsilon \sim Q_s^4$ and $\Delta z \sim 1/Q_s$. 

For one-dimensional random walk in the topological number space 
the topological charge (winding number) generated 
during the time $\tau_B$, when the magnetic field is present, may be
estimated as
\begin{equation}
\label{def1}
n_w \equiv \sqrt{Q_s^2}=\sqrt{ \Gamma_{S} \cdot V \cdot \tau_B} \sim 
\sqrt{\frac{dN_{\rm hadrons}}{dy}} \cdot \sqrt {Q_s \ \tau_B}~,
\end{equation}
where $\Gamma_S$ is the sphaleron transition rate which in 
 weak and strong 
coupling $\Gamma_S \sim  T^4$ with different coefficients. 
The initial temperature $T_0$ of the produced matter at time
$\tau \simeq 1/Q_s$ is proportional to the saturation momentum
$Q_s$, $T_0 = c\ Q_s$. At the last step of (\ref{def1}) the expansion time  
and the corresponding time dependence of the temperature are neglected.
Since sizable  sphaleron transitions occur only in the deconfined phase,
the time $\tau_B$ in Eq.~(\ref{def1}) is really the smallest
lifetime between the strong magnetic field $\tilde{\tau}_B$ one and the
lifetime of deconfined matter $\tau_{\epsilon}$:
\begin{equation}
\tau_B = {\rm min}\{\tilde{\tau}_B, \tau_{\epsilon} \}.
\end{equation}

The measured electric charge particle asymmetry  is associated with the averaged
correlator $a$ by the following relation~\cite{Vol05}:
\begin{eqnarray}
\label{cos}
\langle \cos (\psi_\alpha+\psi_\beta-2\Psi_{RP}) \rangle =  
 \langle \cos (\psi_\alpha+\psi_\beta-2\psi_c) \rangle / v_{2,c}=
v_{1,\alpha} v_{1,\beta} - a_\alpha a_\beta~,
\end{eqnarray}
where $\Psi_{RP}$ is the azimuthal angle of the reaction plane defined by
the beam axis and the line joining the centers of colliding nuclei.
Averaging in (\ref{cos})  is carried out over the whole event ensemble.
The second equality in (\ref{cos}) corresponds to azimuthal measurements
with respect to particle of type $c$ extracted from three-body correlation
analysis~\cite{Vol05}, $v_1$ and $v_2$
are the directed and elliptic flow parameters, respectively.  According 
to Ref.~\cite{Kharzeev:2004ey} an average correlator
 $a=\sqrt{a_\alpha a_\beta}$
is related to the topological charge, $n_w$, as
\begin{equation}\label{rel}
a \sim \frac{n_w}{dN_{\rm hadrons}/dy} \sim 
\frac{\sqrt {Q_s \tau_B}}{\sqrt{dN_{\rm hadrons}/dy}}
\sim \sqrt{\frac{\tau_B}{Q_s}} \sim (\sqrt{s_{NN}})^{-1/16}
\cdot \sqrt{\tau_B},
\end{equation}
where absorption and rescattering in dense matter
responsible  are neglected for the
same and opposite charge correlations. In the last equality we assumed
that $Q_s^2 \sim s_{NN}^{1/8}\sim dN_{\rm
hadrons}/dy$~\cite{KN01}. Our susequent consideration 
is based on  Eq. (\ref{rel}).

\begin{figure}[h]
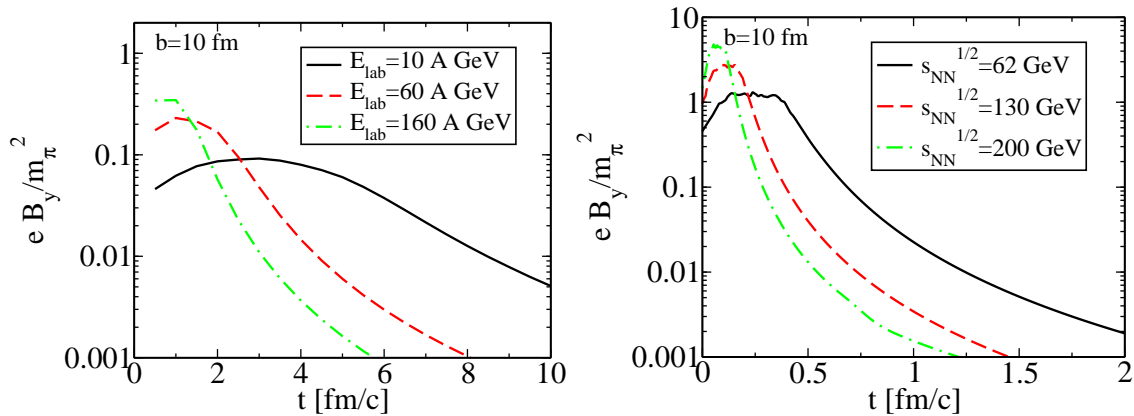

  \centering
\includegraphics[width=0.45\textwidth]{B_semi_log.eps} 
\includegraphics[width=0.45\textwidth]{B_semi_log_RHIC.eps}
\caption{The time evolution of the magnetic field strength $eB_y$
at the central region in  Au+Au collisions  with the impact
parameter $b=10$ fm  for different bombarding energies.
Calculations are carried out within the UrQMD
 model~\cite{Bass:1998ca} (for a detail see~\cite{SIT09}).
   \label{B_ev}}
\end{figure}

\begin{figure}[thb]
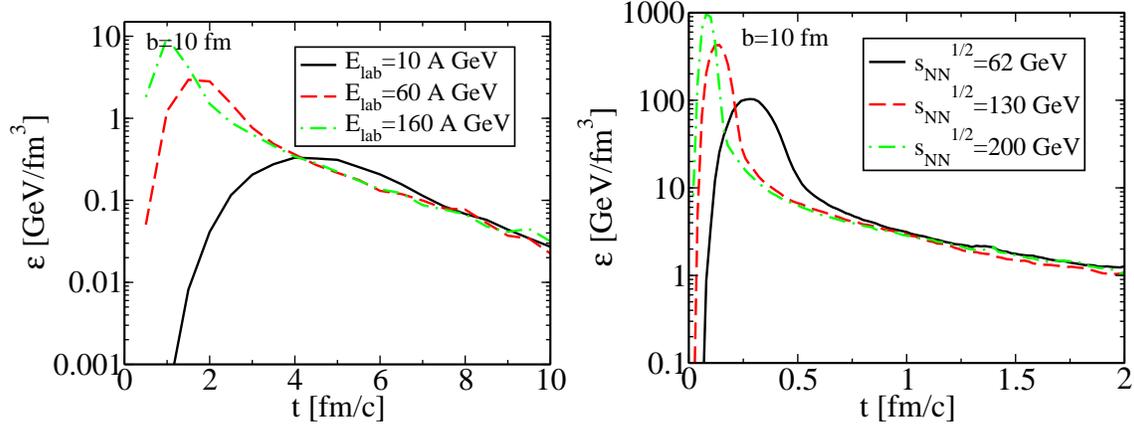

\begin{center}
\includegraphics[width=0.45\textwidth]{E_semi_log.eps}  
\includegraphics[width=0.45\textwidth] {E_semi_log_RHIC.eps}
\caption{ The time evolution of the energy density $\varepsilon$ of created
particles in the Lorentz-contracted box with the 2 fm side
at the central point of overlapping region. The impact parameter $b=10$ fm.
 \label{E_ev}}
\end{center}\end{figure}

Thus, the direct energy dependence of average correlator is 
comparatively weak. Results of
dynamical heavy-ion calculations of the magnetic field at the central
point of the transverse overlapping region of colliding nuclei and energy
density of created particles are
presented in Figs.~\ref{B_ev} and \ref{E_ev}, respectively. Here for a
field estimate we follow Ref.~\cite{SIT09} basing on
the UrQMD model~\cite{Bass:1998ca} and applying
the electromagnetic Lienard-Wiechert potential with the retardation
condition for the magnetic field.
As is seen,  at the impact parameter $b=10$ fm
the maximal strength of the dominant magnetic field component $B_y$
(being perpendicular to the reaction plane) is decreased in Au+Au 
collisions by the factor of about 10, when one proceeds
from $\sqrt{s_{NN}}=$200 GeV to $E_{lab}=$11 GeV, while for  the  created
particle energy density $\varepsilon$ in the central box this factor is 250,
{\em i.e.} noticeably higher.

To use Eq. (\ref{rel}) we need to identify the impact
parameter, saturation momentum and multiplicity at a specific centrality.
These can be found in Ref.~\cite{KN01} where the Glauber calculations were
done. As a reference point we choose $b=$10 fm  in our subsequent
consideration.

The measured value of $\langle \cos
(\psi_\alpha+\psi_\beta-2\Psi_{RP})\rangle$ for the same charge particles
from   Au+Au ($\sqrt {s_{NN}}=200$ GeV) collisions at the impact parameter
$b=$10 fm (40-50$\%$ centrality interval)  is
$-(0.312\pm 0.027)\cdot 10^{-3}$~\cite{Vo09}.
Appropriate number for $\sqrt {s_{NN}}=62$ GeV seems to be a
little bit larger but for Cu+Cu collisions the effect is
definitely stronger~\cite{Vo09}.
Thus, ignoring any final state interactions with medium, assuming $a_\alpha
=a_\beta=a$ and neglecting the directed flow
$v_{1a}=v_{1b}=0$ we get from Eq.~(\ref{cos}) $a^2_{exp}=0.31\cdot
10^{-3}$ for the maximal RHIC energy. Using numbers for the  
$\sqrt{s_{NN}}=$200 GeV reference case, from Eq.~(\ref{rel}) 
we may quantify the $\cal{CP}$ violation effect  by the correlator
\begin{equation}
\label{res3}
a^2  =  K_{Au} \  (\sqrt{s_{NN}})^{-1/8} \cdot \tau_B~.
\end{equation}
The normalization constant $K_{Au}$ can be tunned  at the reference energy
$\sqrt{s_{NN}}=$200 GeV from the inverse relation and experimental
value $a_{exp}$ at this energy for $b=$10 fm
\begin{equation}
\label{res4}
K_{Au}=\frac{ a^2_{exp} \cdot (200)^{1/8} } { \tau_B(200)}~.
\end{equation}
The lifetime $\tau_B$ may be defined as the time during which the
magnetic field is above the critical value needed to support a
fermion Landau level on the domain wall $eB_{crit} = 2 \pi/ S_d$,
where $S_d$ is the domain wall area. Since the size of the domain
wall is not reliably known, it is hard to pin down the number, but
it should be of the order of $m_\pi^2$. Honestly, we have to treat
it as a free parameter.

Indeed the size of the topological defect (say, a sphaleron) in the
region between $T_c$ and 2$T_c$
is very uncertain. At weak coupling, the size is determined by the
magnetic screening mass and it  is $\sim 1/(\alpha_s T)$.
If one plugs $\alpha_s\approx$0.5 and $T =$ 200 MeV, the size is  of
about 2 fm and then the threshold field
is very small $eB_y \sim (\alpha_s T)^2 \sim 0.2 \ m_\pi^2$.

On the other hand, we know that between $T_c$ and $2T_c$ the magnetic
screening mass which determines the size
of the sphaleron is not small as expected from the perturbative theory,
$\alpha_s T$, but from the lattice it is  numerically large till about
5$T_c$. This would increase the
threshold to 20 $m_\pi^2$, however the relation between magnetic mass and
the sphaleron size is valid only as long as the coupling is weak.
All we can say it is perhaps in between (0.2$-$20) $m_\pi^2$. Eventually
lattice QCD calculations may clear this up.

\begin{table}
\caption{
Estimated parameters for the ${\cal CP}$ violation effect in Au+Au
collisions at centrality (40-50)$\%$ with the critical field
$eB_{crit}=0.2 \ m_\pi^2$.
}
\label{tabl2}
\begin{center}\begin{tabular}{|c|c|c|c|c|}
\hline\noalign{\smallskip}
$\sqrt{s_{NN}} \ $GeV & \ $s_{NN}^{1/16}$ \ & $\tilde{\tau}_B$, fm/c& $\tau_\epsilon$, fm/c & $a^2$ \\
\hline\noalign{\smallskip}
$4.5 \cdot 10^3$ & 2.86  & 0.018  & $>$1 &0.016$\cdot 10^{-4}$  \\
200    & 1.94 & 0.24 & $>$2 &0.31$\cdot 10^{-3}$  \\
130    & 1.84  & 0.33& $\sim$2.3 &0.45$\cdot 10^{-3}$ \\
62     & 1.68  & 0.62 & $\sim$2.2  &0.93$\cdot 10^{-3}$ \\
17.9   & 1.43 & 1.41 & $\sim$2. & 2.48$ \cdot 10^{-3}$\\
11. & 1.35 & 1.66&$\sim$ 1.9 &3.10$\cdot 10^{-3}$\\
4.7  & 1.21 & 0. & 0. &0.\\
\hline\noalign{\smallskip}
\end{tabular} \end{center}
\end{table}

The upper bound on the magnetic strength $eB_{crit}=20\ m_\pi^2$
results in $\tau_B=0$  even for the RHIC energy and therefore in this
case the CME should not be observable at all in this energy
range. The time evolution of the magnetic field and energy density,
$\varepsilon$, of newly created hadrons are presented in
Figs.~\ref{B_ev} and \ref{E_ev}. The extracted values of $\tau_B$
defined by the constraints $eB_y>0.2\ m_\pi^2$  and
$\tau_\epsilon$ ($\epsilon >1 \ {\rm GeV} /{\rm fm}^3$) are summed
in Tabl.~\ref{tabl2}. For the reference energy and the  minimal
magnetic field constraint we have
$K_{Au}=2.52\cdot 10^{-3}$.
If lifetimes are known for all energies one can estimate the
$\cal{CP}$ violation effect through the $a^2$ excitation function.

From the first glimpse as follows from Tabl.~\ref{tabl2}, in the case of
$eB_{crit}=0.2 \ m_\pi^2$ the interaction time $\tau_B$
is defined solely by evolution of the magnetic field since
$\tilde{\tau}_B<\tau_\varepsilon$ whereas $\tau_\varepsilon \approx 2$ fm
independent of $\sqrt {s_{NN}}$. The expected CME for Au+Au at $b=10$ fm
(see the last column in Tabl.~\ref{tabl2}) monotonously increases when
$\sqrt{s_{NN}}$ goes down but then sharply vanishes exhibiting
a shallow maximum in the range between near the top SPS and NICA
energies. The position of CME maximum and its magnitude depend on the cut
level which just defines $\tilde{\tau}_B$.
The decrease of the $eB_y$ bound till 0.02\ $m_\pi^2$ shifts the
maximum toward lower  energy $\sqrt{s_{NN}}$ and enhances
its magnitude. In an opposite limit
when  results are extrapolated to the LHC energy, the CME falls down by a
factor of about 20 with respect to the RHIC energy. This result is quite
understandable. The CME is mainly defined by the relaxation time of the
magnetic field which is concentrated in the Lorentz-contracted nuclear
region $\sim 2R/\gamma$. Therefore, the CME is inversely proportional to the
colliding energy, $\sim 1/\sqrt{s_{NN}}$, and proceeding from the RHIC to
LHC energy we roughly get the suppression factor about 4.5/0.2$\approx$ 22.

There is one worrying point here. Proceeding from  $\sqrt{s_{NN}}=$200
to 62 GeV the predicted value of $a^2$ for $b$=10 fm increases in three
times though not more 20$\%$ growth has been observed in these collisions
in the recent experiment~\cite{Vo09}. This essential disagreement cannot
be removed by a simple variation of $eB_{crit}$. One may try to explain
this correlator overestimation at $\sqrt{s_{NN}}=$62 GeV by an irrelevant
choice of the energy dependence of multiplicity in Eq.~(\ref{res3}). For
the correlator ratio  at these two energies we have
\begin{eqnarray}
\frac{a^2(200)}{a^2(62)}=\frac{\tau_B(200)}{\tau_B(62)}
\left( \frac{62}{200}
\right)^{1/8} =0.387 \ (0.31)^\beta \approx 0.72.
\label{ratio}
\end{eqnarray}
where we use lifetime values from Tabl.~\ref{tabl2} and experimental values
for correlators~\cite{Vo09}, $\beta\equiv 1/8$. As follows from
Eq.~(\ref{ratio}), to explain the experiment the exponent should be
negative, $\beta <0$. Therefore, the fast growth of $\tau_B$ with the
energy decrease cannot be compensated by uncertainty in the energy 
dependence of the correlator $a$.

\begin{figure}[h]
\centering\includegraphics[angle=-90,width=0.6\textwidth] {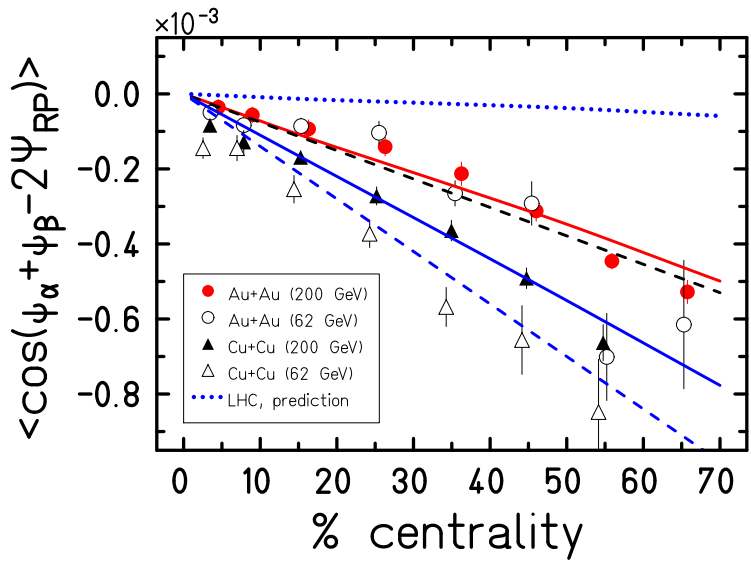} 
\caption{Centrality dependence of the CME. Experimental points for Au$+$Au
and Cu$+$Cu collisions are from~\cite{Vo09}. The dotted line is our
prediction for Au+Au collisions at the LHC energy.
 \label{CME_cu}}  
\end{figure}

Uncertainty in the choice of the impact parameter does not help us to 
solve this issue. It turned out that one fails to fit this ratio by the 
variation of only $eB_{crit}$. Here we should remember that not only the 
strong magnetic field but also high density of soft equilibrium 
quark-gluon matter are needed. Equilibration requires some finite 
initial time $t_{i,\varepsilon}$
which we associate with the moment when a maximum in the $\varepsilon(t)$
is achieved (see Fig.\ref{E_ev}). This makes $\tilde\tau$ shorter and in 
combination with $eB_{crit}$ variation, 
$\tau_B=\tilde\tau_B-t_{i,\varepsilon}$, allows us to satisfy 
the condition (\ref{ratio}). Using the value of $\tau_B(200)$
obtained in this analysis one can recalculate the coefficient in
Eq.~(\ref{res3}), $K_{Au}=6.05\cdot 10^{-3}$, and therefore find the
correlator $a$ at any energy. In principle, similar analysis may
be repeated for other impact parameters to consider the
$b$-dependence of the CME. As was shown in
Refs.~\cite{KMcLW07} the CME roughly is linear in $b/R$.
Taking this as a hypothesis we evaluate the centrality dependence
of the CME fitting this line to points $b=10$ fm (or centrality
$(40-50)\%$) to be estimated in our model and $b=0$ where the CME
is zero. The results are presented in Fig.~\ref{CME_cu} for Au+Au
collisions at three energies.

As it is seen the calculated lines quite reasonably reproduce the
measured points of azimuthal asymmetry of charge separation for
Au$+$Au collisions at $\sqrt{s_{NN}}=$200 and 62 GeV. The chosen value
of $eB_{crit}=0.7\ m_\pi^2$ results in absence of the CME
at the top SPS energy because  the critical magnetic field practically
coincides with the maximal field at this bombarding energy (see
Fig.~\ref{B_ev}). The CME at the LHC energy is expected to be
less by a factor of about 20  as compared to that at the RHIC energy.
Note that at the LHC energy we applied a simplified
semi-analytical model \cite{SIT09} for magnetic field creation and
assumed $t_{i,\varepsilon}=0$. Thus, we consider this LHC estimate
as an upper limit for the CME.

Similar analysis can be repeated for Cu+Cu collisions basing
on available RHIC measurements at two collision energies.
Here one remark is in order.
An enhancement of the CME in Cu+Cu collisions with respect to Au+Au ones
was seen experimentally at the same {\em centrality}~\cite{Vo09} but not 
at the same {\em impact parameter}. As follows from the Glauber
calculations, the impact parameter b=10 fm for gold reactions
corresponds to centrality (40-50)$\%$  while the same centrality for copper
collisions  matches  b=4.2 fm. The
time distributions of the magnetic field and energy density for Cu+Cu
collisions look very similar to that for Au+Au ones but  lifetimes, both
$\tilde{\tau}_B$ and $\tau_\varepsilon$, are shorter  in the Cu+Cu case.
For the extracted lifetimes and other characteristics at $eB_{crit}=0.2
m_\pi^2$ ($K_{Cu}=6.34\cdot 10^{-3}$) 
we meet again the same problem: one should compensate a too
strong energy dependence of the model correlators by the proper definition
of lifetimes. Defining the lifetime  in the same manner as for Au+Au
collisions  the lifetime ratio $\tau_B(62)/\tau_B(200)$
is turned out to be very close to experimental one at  $eB_{crit}=$0.3 
$ m_\pi^2$. In this case
$K_{Cu}=11.9\cdot 10^{-3}$. In the linear approximation
with the reference point at $b=$4.2 fm, one may draw
the centrality dependence of the CME for Cu+Cu collisions shown also
in Fig.\ref{CME_cu} which is in a reasonable agreement with the experiment.
Note that $eB_{crit}=$0.3 $ m_\pi^2$ which is slightly above the maximal
magnetic field at $\sqrt{s_{NN}}=$62 GeV implies that the CME for Cu$+$Cu
collisions will not be observable even at the top SPS energy.

From dimensionality arguments the system-size dependence of  the chiral
magnetic effect  (at the same all other conditions) would be expected to be
defined by the surface $S\equiv S_{\rm A}(b)$ of an ``almond''-like
transverse area of overlapping nuclei
since  both the high magnetic field and deconfined
matter are needed for this effect. The magnetic field was evaluated
in the the center of the overlapping  region but as was shown
in Ref.~\cite{SIT09} the studied  $eB_y$ component is quite homogeneous
along $x$ of this ``almond''.  
Using  for ``almond'' area  a  rough estimate as two
overlapping discs of radius $R=r_0 A^{1/3}$, namely
$S\equiv S_A(b)=\pi \sqrt{R^2 - (b/2)^2} (R-b/2)$,  we have
$S_{\rm Cu}(b=4.2)/S_{\rm Au}(b=10)\approx$ 1.65 which seems to be
consistent with experimental ratio of
the CME at $\sqrt{s_{NN}}$ for these two points. However, this result was 
obtained for different $eB_{crit}$ and non-zero initial
time $t_{i,\varepsilon}$,  and this success cannot be repeated
for Cu+Cu (62 GeV) collisions. Therefore,
the Cu enhancement effect is not only a geometric one.

\section{Towards a kinetic approach to the CME background}
\label{sec:3}
The discussed CME signal, the electric charge
asymmetry with respect to the reaction plane,  may originate not
only from the spontaneous local {\cal CP} violation but also be
simulated by other possible effects. In this respect it is important
to consider the CME background. We shall do that considering a full
evolution of  nucleus-nucleus collisions in terms of the HSD transport
model \cite{HSD} but including formation of electromagnetic field as
well as its evolution and impact on particle propagation.

Generalized on-shell transport equations for strongly interacting particles
in the presence of magnetic fields can be written as

 \be \label{kinEq}
 \{ \frac{\prt}{\prt t}+\left(\nabla_{\vec{p}} \ \vec{U}\right)
 \nabla_{\vec{r}}-\left(\nabla_{\vec{r}} \ \vec{U}+{ q\vec{v}\times
 \vec B} \right)\nabla_{\vec{p}} \
 \} \ f(\vec{r},\vec{p},t)\nonumber =I_{coll}(f,f_1,...f_N)
\ee
 which are supplemented by the wave equation for the magnetic field whose
solution in the semi-classical approximation for point-like moving charges
is reduced to the retarded Li\'enard-Wiechert  potential used 
above~\cite{SIT09}.  The term $U\sim Re (\Sigma^{ret})/2p_0$ is
 the hadronic mean-field. 
 
 \begin{figure}[thb]
\centering
\includegraphics[width=0.45\textwidth]{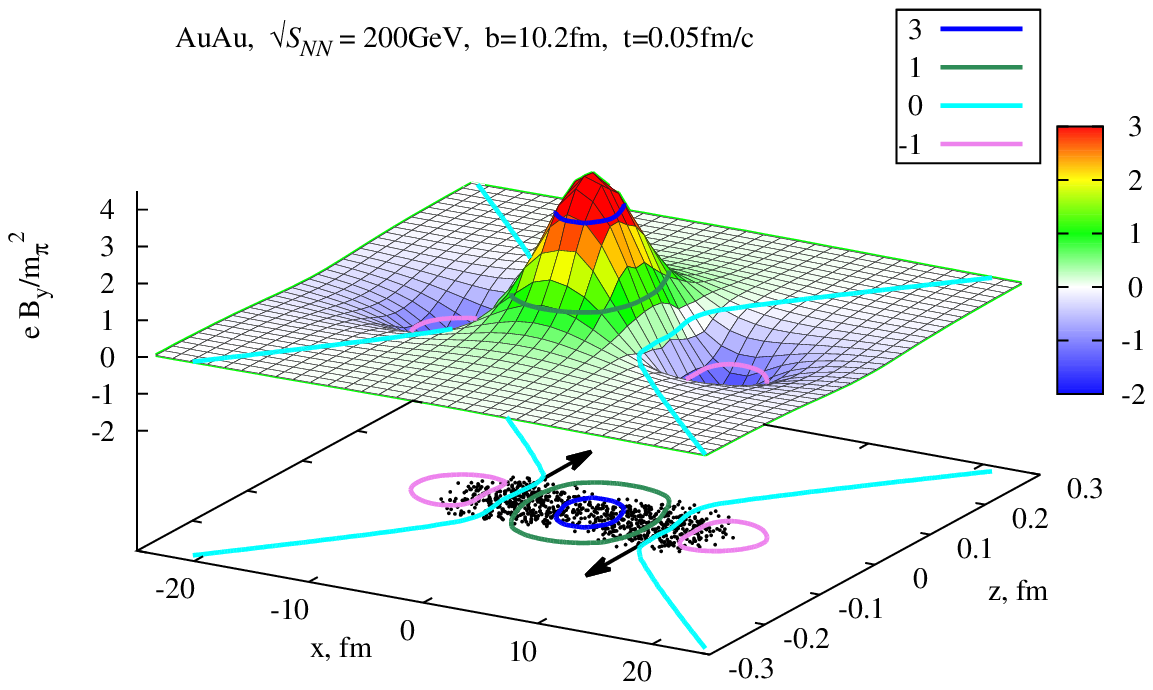} 
\includegraphics[width=0.45\textwidth]{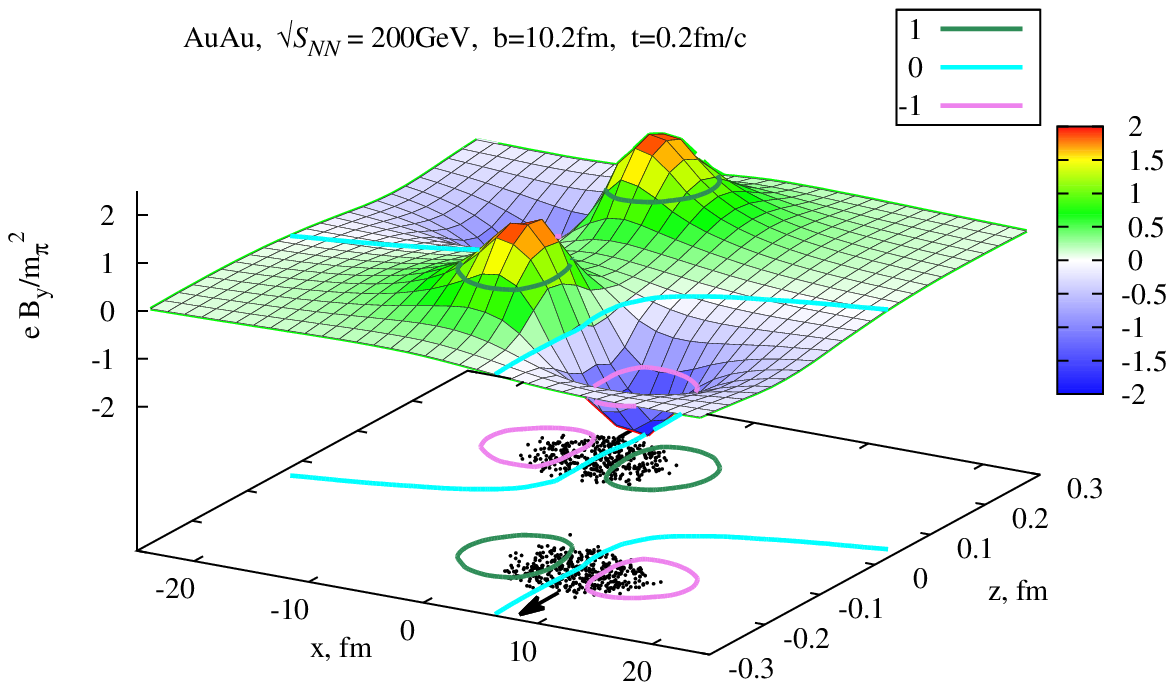}  
\caption{Distribution of the magnetic field strength $eB_y$
in the $y=0$ plane
at $t=$0.05 and 0.2 (in the middle) fm/c for Au+Au collisions
at $\sqrt{s_{NN}}=$200 and $b=$10.2 \ fm. The collision geometry is
projected on $x-z$ plane by points corresponding to a particular
spectator position.
Curves (and their projections) are levels of the constant $eB_y$.
  \label{By} }
\end{figure}

One should note that the off-shell HSD transport approach 
is based not on the Boltzmann-like
transport equation (\ref{kinEq}) but rather on the off-shell Kadanoff-Baym
equations~\cite{CJ99} having similar general structure. The set of
equations was solved in a quasiparticle approximation by using the
Monte-Carlo parallel ensemble  method.
To find the magnetic field a space grid was used.
In a lattice point of this grid  the retarded vector
potential is evaluated. The magnetic field is calculated by its 
numerical differentiation.
The field inside a cell is approximated by that at
the nearest grid point. To avoid singularities and
self-interaction effects, particles within a given
cell are excluded from  procedure
of the field calculation.

An evolution snapshot of the magnetic field $B_y(x,y=0,z,t)$ (in units
of $m_\pi^2$) formed in Au+Au (200 GeV) peripheral ($b=$10.2 fm) collisions
are given in Fig.\ref{By} for two time moments $t=$0.05 and 0.20 fm/c. 
The collisional geometry is presented by a set of points every of
which corresponds to a spectator nucleon. The whole field is not homogeneous
exhibiting a wide maximum over the transverse size of overlapping
(participant) matter and strong contraction in longitudinal direction. 
Opposite rotation of the magnetic field along
direction of two colliding nuclei results in corresponding two minima
from outer sides of spectator matter remnants. At expansion these remnants
are moving away from each other.  The position of a maximum in the magnetic 
field strongly correlates with that in the
energy density  of created particles.  Large local values of $B_y$ 
and $\varepsilon$ reached
in these Au+Au collisions provide necessary conditions for  observation of
signals of a possible parity violation.

\section{Discussion and conclusions}
Summarizing  one should note that for
heavy-ion collisions at
$\sqrt{s_{NN}}\gsim$ 11 GeV the magnetic field and energy density
of deconfined matter reach very high values which seem to be high
enough for manifestation of the Chiral Magnetic Effect. However,
these are only necessary conditions. To estimate a possible CME
a particular model is needed.  For the average correlator  our
qualitative prediction $a^2\sim s^{-1/8}_{NN}$  has
a rather small exponent but nevertheless it is too strong to
describe observable energy behavior of
the CME. This model energy dependence can be reconcile with
experiment~\cite{Vo09} by a detailed treatment of the lifetime
taking into account both the time of being in a strong magnetic field
and time evolution of the energy density in the QGP phase. For the
chosen parameters we are able to describe data for Au+Au collisions on
electric charge separation at two available energies. We predict that 
the effect
will be much smaller at the LHC energy and will sharply disappear near
the top energy of SPS. Coming experiments at the Large Hadron
Collider and that the planned Beam Energy Scan program  at RHIC  
\cite{BES} are of great interest since they will allow one to test the
CME scenario and to infer the critical magnetic field
$eB_{crit}$.

The experimentally observed CME enhancement for Cu+Cu collisions
is related with the selection of different impact parameters for
the same centrality. However,
it is not reduced to a purely geometrical effect.

The problem of parity violation in strong interactions and the related
CME are actively debated now.  It is of great
interest that the electric charge
asymmetry with respect to the reaction plane  may originate not
only from the spontaneous local {\cal CP} violation but also be
simulated by other possible effects. First step in study of dynamical
study of the CME background
 has been made in Sec.\ref{sec:3}. 
It is important that the developed kinetic approach in principle allows
one to simulate the
Chiral Magnetic effect itself. This work is in progress.

\section*{Acknowledgements}
Successful collaboration with D. Kharzeev, V. Skokov, E. Bratkovskaya,
W. Cassing, V. Konchakovski and S. Voloshin is greately acknowledged.
We are thankful to V. Koch, R. Lacey, I. Selyuzhenkov, O. Teryaev  and
J. Thomas  for comments.  V.T.  is partially
supported by the DFG grant WA 431 8-1 RUSS
and the Heisenberg-Landau grant. 
V.V. acknowledges financial support within the ``HIC for FAIR" center
of the ``LOEWE'' program.

\end{document}